# Assessment of East-West (E-W) and South-North (S-N) facing Vertical Bifacial Photovoltaic Modules for Agrivoltaics and Dual-Land Use Applications in India

Nishant Kumar[a,#], Shravan Kumar Singh[a], Nikhil Chander[a,b*]

[a] Department of Electrical Engineering, Indian Institute of Technology Bhilai, Durg-491002, Chhattisgarh, INDIA

[b] Department of Electronics and Communication Engineering, Indian Institute of Technology Bhilai, Durg-491002, Chhattisgarh, INDIA

**Abstract**

Deploying vertical bifacial PV modules can play a significant role in agrivoltaics, fencing walls, noise barriers, building integrated photovoltaics (BIPV), solar PV for electric vehicles, and many other applications. This research work presents the performance comparison of vertical bifacial photovoltaic (VBPV) modules facing East-West (E-W) and South-North (S-N) directions. Also, the VBPV modules are compared with vertical and tilted south-facing monofacial PV modules. Six PV modules (monofacial and bifacial) were installed at the rooftop of IIT Bhilai academic building, Raipur (21.16° N, 81.65° E), India, and studied for a year from May 2022 to April 2023. The results show that the E-W facing VBPV module gives two production peaks, one in the morning and another in the evening, as compared to the single notable rise at midday observed for a monofacial module. From a series of experiments, 19 days of data were collected over the one-year period from May 2022 to April 2023, with specific inclusion of important days like solstices and equinoxes. In addition, the energy generation results are compared with PVsyst simulations, while also addressing the limitations of the PVsyst simulation of vertical PV modules. E-W bifacial generation is higher than S-N bifacial and south-facing monofacial modules from February to April. The VBPV modules in E-W and S-N orientations present a promising opportunity for expanding the agrivoltaics sector in tropical and sub-tropical countries, like India. This has huge implications for addressing the sustainable development goals by simultaneously contributing to sustainable land management, green energy generation, energy security and water conservation in the vast geo-climatic expanse of tropics.

# Nishant Kumar is presently pursuing PhD at National Centre for Photovoltaic Research and Education, Indian Institute of Technology Bombay, Mumbai-400076, INDIA

## 1. Introduction

The installation of solar photovoltaic (PV) plants has increased exponentially in the last few decades. This is happening due to several considerations such as electricity cost reduction, environmental concerns ($CO_2$ emissions), and a finite supply of non-renewable resources. India has enormous solar energy potential. Approximately 5,000 trillion kWh of energy is incident over India's land area each year, with most areas getting 4-7 kWh per square meter per day [1]. However, utilizing this solar energy is very difficult due to location dependency, low efficiency of solar cells, improper installation of existing solar PV modules and lack of technological innovations in installing PV modules. According to the Installed Capacity Report [2] by Central Electricity Authority (CEA), Government of India, India has an installed capacity of 184 GW from renewables, excluding hydro-power, as on 31st June, 2025, which is approximately 38% of the total installed power capacity of 484 GW. Solar photovoltaics (SPV) is the major contributor to this figure, with 116 GW of cumulative installations in India. Monofacial PV modules have been traditionally used for ground-mounted and solar rooftop projects. However, with the advent of bifacial PV modules, monofacial modules are increasingly becoming a thing of the past.

Bifacial solar PV (BPV) modules have some special features and advantages over monofacial solar PV modules. Bifacial PV technology requires lesser land area in vertical orientation. It makes dual use of the available land area, farming and harvesting solar energy to produce electricity on the same land. According to the International Technology Roadmap for Photovoltaics (ITRPV) [3], bifacial PV cells made up around 20% of the global PV cell market in 2020 and are anticipated to reach 70% of the market by 2030. In India, the Delhi Metro Rail Corporation Limited (DMRC), installed 100 kWp vertical bifacial PV modules along its elevated corridor between Jamia Millia Islamia and Okhla Vihar section of the magenta line [4]. Deployment of vertical bifacial modules can play a significant role in agrivoltaics, which combines farming and generation of electricity on the same land, implying land use for dual purposes. Bifacial solar PV modules aren't limited to agrivoltaic applications. Apart from this, the bifacial PV modules can be deployed as fencing walls, railings, building integrated photovoltaics (BIPV) and many other current and future applications.

Researchers have studied the effectiveness and the way of installation of bifacial PV plants at various locations and climatic conditions to extract maximum energy throughout the day, and simultaneously minimize the land usage. Next2Sun [5] has investigated and documented bifacial PV agrivoltaic systems of 33 kWp (supply for 11 households) put into operation in Sweden in the year 2021, South Korea's first agrivoltaics ground-mounted plant of 30 kWp (two plants) put into operation in the year 2020, Europe's first bifacial ground-mounted photovoltaic system of 2 MWp (Supply for 700 households) put into operation in the year 2018. Baumann et al. [6] studied the ground coverage ratio

(GCR) and albedo of a bifacial PV plant having an installed capacity of 9.09 kWp located in Winterthur (coordinates: 47°30′ N 8°43′ E), Switzerland. Jouttijärvi et al. [7] demonstrated that vertical BPV can be especially helpful at high latitudes due to the low average solar altitude angle, which allows vertical surfaces to efficiently gather irradiation for several hours. Low solar altitude angle improves light collection on vertical surfaces, or in other words, low solar altitude angle enhances light collection on vertical surfaces. India and other sunny countries have already canceled plans in 2017 for massive coal power plants in favor of solar PV systems as solar PV energy prices are lowering day by day [8]. The performance assessment of bifacial PV in article [9] showed soiling rates of 0.328%/day (front) and 0.031%/day (rear), with bifacial gain of 13.98% (soiled) versus 12.92% (clean). Vertical installation reduced soiling losses to ~2%. The effect of wind speed and relative humidity were also studied in this research. In the article [10], authors presented module technology for agrivoltaics and comparison with the conventional S-N tilted monofacial and bifacial agrivoltaics (AV) farms. A rigorous analytical framework has been developed to precisely calculate the sunlight intercepted by elevated PV arrays and the transmitted photosynthetically active radiation (PAR) below the elevated PV array. The energy generation efficiency of bifacial modules is determined by both the front-side and the rear-side output power. The authors in [11] compared the I-V parameters as well as field performance of different module structures to measure energy yield of bifacial modules. Their results showed that the glass/glass bifacial modules encapsulated with bifacial solar cells provided over 6% more energy yield compared to the glass/ backsheet monofacial modules encapsulated with regular monofacial solar cells. In another paper [12], researchers compared the performances of the two bifacial PV system orientations, optimally tilted facing south/north, and vertically installed facing east/west under no-shading and shading conditions by using RADIANCE ray-tracing software.

Literature review reveals that many reported experiments study standalone systems that over-represent the yield performance obtainable in farms. Researchers analyzed the fixed-tilted bifacial farm configurations, namely south-facing monofacial, south-facing tilted bifacial (TBF), and ground-sculpted vertical bifacial (VBF) arrays, at Dhaka, Bangladesh (23.7 ◦N, 90.4 ◦E) [13]. The optimal TBF configuration (albedo=0.5) yields 21.3% and 73.3% more than the optimal monofacial and the optimal VBF configurations, respectively. Similarly vertical configuration was explored for a UK case study. Another study [14] reveals that a vertically mounted bifacial PV (VBPV) system produces much more energy than both the vertically mounted monofacial PV (VMPV) and the conventional tilted monofacial PV (TMPV) systems. It generates 7.12%–10.12% more power than the VMPV system daily and 22.88%–26.91% more than the TMPV system in the early morning and late afternoon. Seasonal analysis reveals higher energy gains, with increases of 11.42% in spring, 8.13% in summer, 10.94% in autumn, and 12.45% in winter compared to the VMPV system. The gains are even greater against the TMPV system, reaching 24.52% in winter. For higher latitude regions, these vertical systems are explored. One of the studies [15] is from an outdoor lab in Trondheim, Norway. It shows the behavior of the systems operating under low sunlight, short sunshine hours, and severe extreme weather. Various BPV setups, including vertical east–west (E–W) and south–north (S–N) orientations, fixed south-tilted panels, and two-axis tracking systems, are evaluated. Results show that mixed orientations improve energy generation by aligning better with consumption needs. Vertical bifacial PV modules respond well to diffused light, especially in vertical setups. Spring conditions enhance performance, though snow reduces efficiency, particularly in tilted configurations. Vertical bifacial make dual land use, e.g., agrivoltaics. Countries struggle to fulfill food demands while simultaneously tackling climate change. However, this technology has advanced the most in Europe, with slower progress in other regions [16]. For that, awareness and clearer policies are needed. As authors in [17] examined the spread of agrivoltaics in Italy, where this innovative solar technology faces multiple challenges. Four key themes emerge-ambiguity, justice, agronomic risk, and exploitation-highlighting the need for clearer policies.

Baghel and Chander [18] studied the performance comparison of monocrystalline and polycrystalline solar PV modules in east-central India. The results showed that the monocrystalline PV module outperforms the polycrystalline PV module with more efficiency, a higher performance ratio (PR), and a better specific yield than polycrystalline PV modules. Vertical bifacial configuration was not explored for East-Central Indian location. Bhaduri and Kottantharayil [19] demonstrated lower soiling rates for vertical mounted BPV modules in Mumbai, India. Raina et al. [20] and Basak et al. [21] studied tilt angle optimization for bifacial PV modules in north-west and southern parts of India, respectively. In the present research work, we look at the performance comparison of vertical bifacial solar photovoltaic (BPV) modules facing E-W, S-N, and fixed-tilted south-facing monofacial, vertical south-facing monofacial at the rooftop of IIT Bhilai, GEC Campus, Sejbahar, Raipur (21.16° N, 81.65° E). The BPV modules are compared with vertical and tilted south-facing monofacial PV modules. The details are provided in the subsequent sections. Further, this research article is beneficial for the PV research community and policy makers to know the performance of bifacial PV modules in vertical orientations. It will be a great help to farmers, engineers, system integrators, and shed light on the way of installation of bifacial PV plants, and use of the land for maximizing farming area for the applications of agrivoltaics in India.

## 2. Materials and methods

In this section, the details of the photovoltaic modules, equipment, location and the experimental setup are provided. Relevant equations and figures are shown and discussed, along with sun-path diagram and shadow analysis, to properly illustrate the experiments and data recording procedure.

## 2.1. Site description and experimental setup

Our experimental set-up consists of six PV modules (monofacial and bifacial) installed on the rooftop of IIT Bhilai, GEC Campus, Sejbahar, Raipur, India (21.16° N, 81.65° E) at a height of 60 feet from the ground surface. The experimental set-up consists of solar PV modules and measuring equipment: I-V tracer, solar survey meter and temperature sensors. Three pyranometers (EKO MS-40) are independently deployed for the experiment. One of the pyranometers measures global horizontal irradiance (GHI), while the remaining two are used to measure the plane of array (POA) irradiance of the front and rear-side for bifacial PV module, specifically for the south-north (S-N) facing vertical bifacial PV module. The measurement of plane of array (POA) irradiance is also carried out by a solar survey meter that is integrated wirelessly with I-V tracer (Seaward) which measures instantaneous irradiance. The outdoor experimental set-up of mono-facial and bifacial solar PV modules and it's nomenclature of the modules are given in **Table 1**.

**Table 1:** Nomenclatures of PV Modules

| Nomenclature of PV Modules | Description |
| --- | --- |
| SF81B | South facing bifacial PV module at 81° tilt (Vertical configuration) |
| EF81B | East facing bifacial PV module at 81° tilt (Vertical configuration) |
| SF81MM | South facing monocrystalline monofacial PV module at 81° tilt (Vertical configuration) |
| SF81PM | South facing polycrystalline monofacial PV module at 81° tilt (Vertical configuration) |
| SF21PM | South facing polycrystalline monofacial PV module at 21° tilt (Conventional Tilt) |
| SF21MM | South facing monocrystalline monofacial PV module at 21° tilt (Conventional Tilt) |

The electrical specifications of the monofacial and bifacial PV modules at standard test condition (STC) are given in **Table 2**. The monofacial PV modules are rated 375 Wp (monocrystalline), 330 Wp (polycrystalline) and the bifacial PV module is rated 355 Wp (n-type PERT).

**Table 2:** Electrical specification of PV modules at STC*

| Datasheet Parameters | Vikram Solar (Monofacial) | | Adani Solar (Bifacial)** |
| --- | --- | --- | --- |
| | Mono SOMERA VSM.72.375.05 | Poly ELDORA VSP.72.330.03.05 | ELAN Series n-type PERT |
| Peak Power Pmax ($W_p$) | 375 | 330 | 355 |
| Number of cells in Module ($N_s$) | 72 | 72 | 72 |
| Maximum Voltage $V_m$ (V) | 40.1 | 38 | 37.9 |
| Maximum Current $I_m$ (A) | 9.36 | 8.7 | 9.37 |
| Open Circuit Voltage $V_{OC}$ (V) | 48.7 | 46.3 | 46.4 |
| Short Circuit Current $I_{SC}$ (A) | 9.94 | 9.24 | 9.74 |
| Module Efficiency η (%) | 19.33 | 17.03 | 17.59 |
| Tc of Open Circuit Voltage (β) | - 0.28%/°C | -0.29%/°C | -0.31%/°C |
| Tc of Short Circuit Current (α) | 0.057%/°C | 0.057%/°C | 0.065%/°C |
| Tc of Power (γ) | -0.39%/°C | -0.38%/°C | -0.4%/°C |

*STC: 1000 W/m² of full solar noon sunshine (irradiance) when the panel and cells are at a standard ambient temperature of 25 °C with a sea level air mass (AM) of 1.5 (1 sun).
** The bifaciality factor is 87%.

The maximum output power of monofacial PV modules is given by the manufacturer at STC, but in the case of bifacial PV modules we are able to extract the power from both the front and rear side. The maximum output power of the bifacial PV modules should account for their ability to generate power from both the front and rear sides [22]. Electrical models of the cells/modules (interconnection of non-identical cells; practical scenario is considered) can be given by the following single diode electrical model of a cell as shown in **Fig.1**.

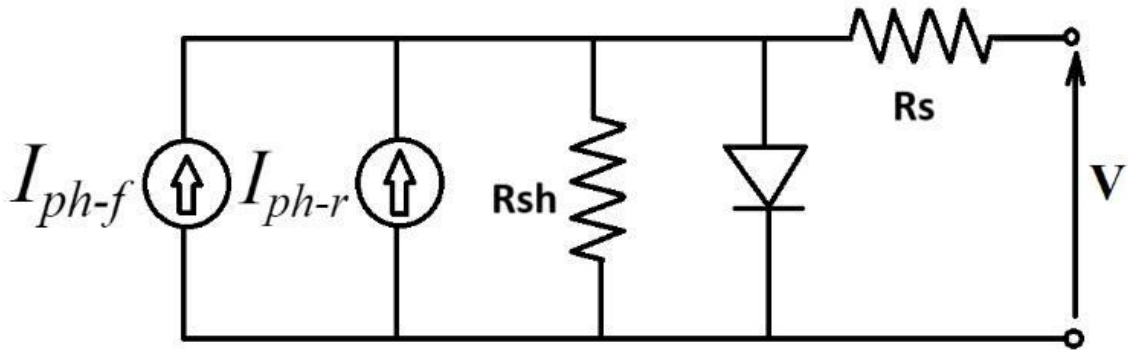

**Fig. 1:** Single diode electrical model of a bifacial solar cell.

Where, $I_{ph\text{-}f}$ = Photocurrent due to front side

$I_{ph\text{-}r}$ = Photocurrent due to rear side

$R_{sh}$ = Parallel resistance

$R_s$ = Equivalent series resistance

The relative performance of the rear side of bifacial modules is described by bifaciality factors (φ) which are defined in IEC TS 60904-1-2 as three ratios. These ratios are determined at STC conditions: specified as 1000 W/m² irradiance level, 25 °C, and an air mass of 1.5, $\varphi P_{max}$, $\varphi V_{OC}$ and $\varphi I_{SC}$ respectively using equation (1,2 and 3).

$$\varphi P_{max} = \frac{P_{max,r}}{P_{max,f}} = \text{Ratio of rear to front side maximum power } (P_{max}) \quad (1)$$

$$\varphi V_{OC} = \frac{V_{oc,r}}{V_{oc,f}} = \text{Ratio of rear to front open-circuit voltage } (V_{OC}) \qquad (2)$$

$$\varphi I_{SC} = \frac{I_{SC,r}}{I_{SC,f}} = \text{Ratio of rear to front short circuit current } (I_{SC}) \qquad (3)$$

The values of $\varphi P_{max}$ typically range from 75% - 95% for n-PERT bifacial modules, and 60% - 70% for p-type PERC bifacial modules. TÜV Rheinland has proposed specific bifacial standard test conditions (BSTC) of 1000 W/m² front-side and 135 W/m² rear-side irradiance [23]. The nominal output power at BSTC of the bifacial module is then measured with an equivalent irradiance ($G_E$) given by equation (4), where bifaciality factor ($\varphi$) is taken into account.

$$G_E = 1000 \text{ W/m}^2 + \varphi * 135 \text{ W/m}^2 \qquad (4)$$

Practically such non-identical cells are interconnected in series or series-parallel combinations to form the modules. Bypass diodes are incorporated into it to avoid the shading or mismatch losses.

## 2.2. Sun Path Diagram

A sun path diagram is a graphical representation of the path of the sun across the sky over a particular location and time period. The diagram typically displays the sun's azimuth and altitude angles at different times of the day and throughout the year. This is also known as a sun chart or sun position chart. The Sun Path diagram is drawn over a year for experiment location by using trial versions of SketchUp and Curic sun tool [https://www.sketchup.com/]. The sun path diagram is shown in **Fig. 2** for the experimental location.

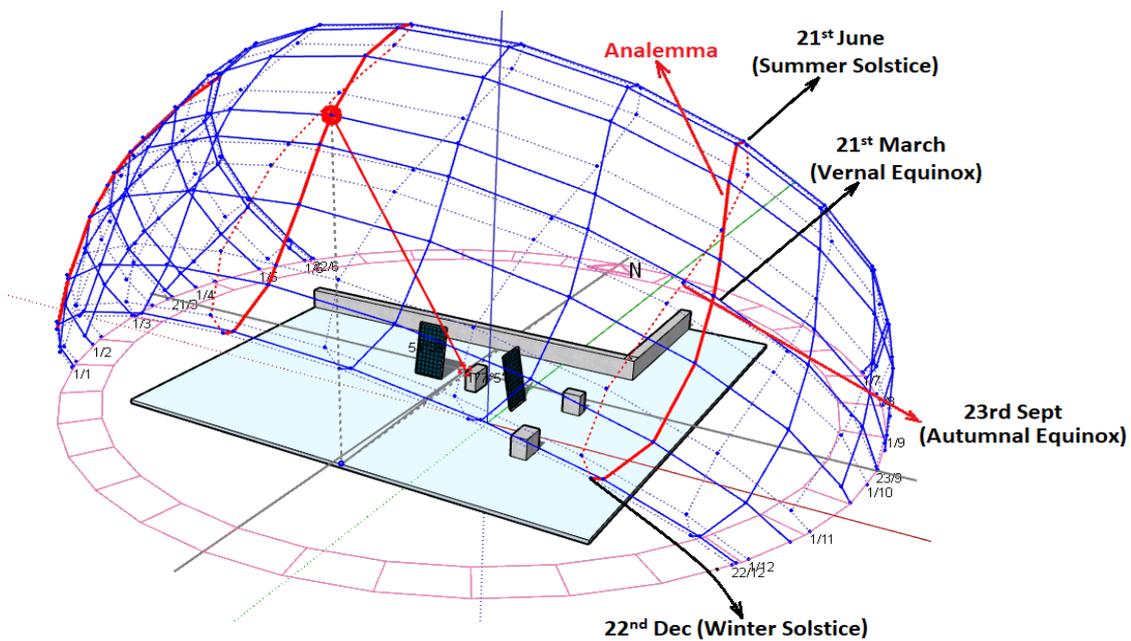

**Fig. 2:** *Sun Path diagram over a year at the experimental location by using Sketchup and Curic sun tool*

## 2.3. Shadow Analysis

Shadow analysis was carried out for every day of the experiment throughout the year. Two extremes (winter and summer solstice) are added as references. According to the shadow analysis, the bottom part of the E-W module is affected by partial shading. Shading is more pronounced a few hours after sunrise and a few hours before sunset. Shadow patterns can be seen throughout the day for summer and winter solstice as a reference. As shown in **Fig. 3**, the bottom part of the E-W facing module is affected by shading. Despite the shadow, we get comparable or more energy generation from the E-W facing module. More details about the energy generation from different configurations are provided in the section 3.

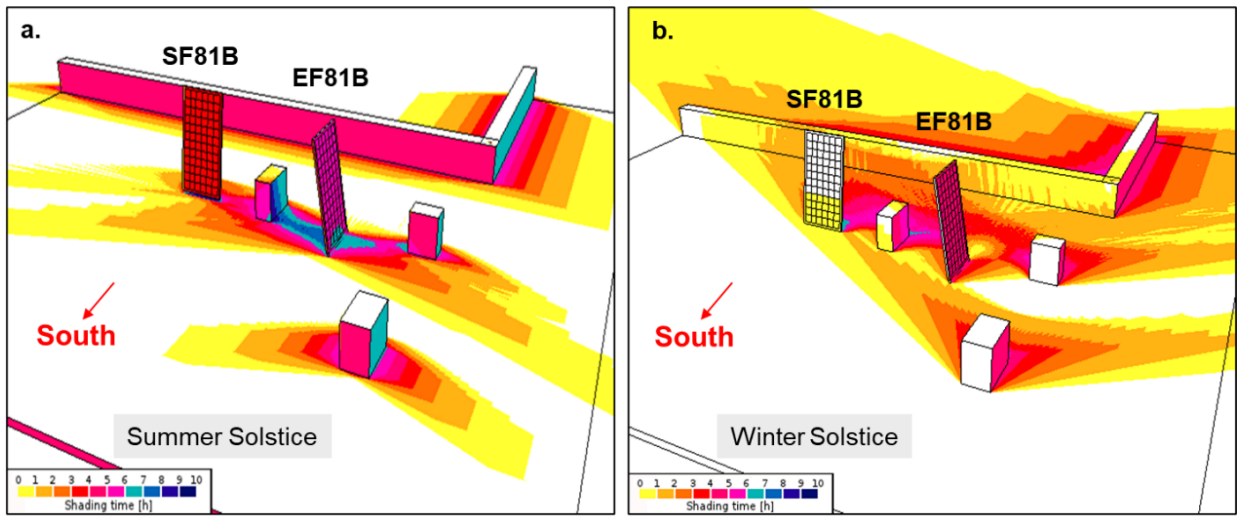

**Fig. 3: a.** *Shadow analysis for summer solstice* **b.** *Shadow analysis for winter solstice*

## 2.4. Thermal Effects on Bifacial Modules

Bifacial panels generally absorb more irradiance than monofacial panels, which can lead to higher operating temperatures. However, their transparent backside allows a significant portion of non-absorbed infrared irradiance to pass through the cell and panel [24]. Based on a basic heat balance model, the operating temperature, ($T_m$) of a solar module can be expressed as

$$T_m = T_a + \frac{\alpha_f G_f + \alpha_r G_r}{U} \qquad (5)$$

Here, $T_a$ represents the ambient temperature, and $G_f$ and $G_r$ denote the irradiance on the front and rear sides, respectively. The absorption coefficients for the front and the rear side are given by $\alpha_f$ and $\alpha_r$, respectively. The absorption coefficients vary for monofacial and bifacial modules. U is the module's effective heat-transfer parameter. The heat-transfer parameter $U$ primarily depends on the type and thickness of the materials used in the module, meaning glass–glass and glass–backsheet modules may have different U values. According to studies by Yusufoglu et al., normal operating cell temperatures of 45 °C were reported for monofacial modules and 47 °C for bifacial modules, suggesting that the specific bifacial modules examined had a lower U value than their monofacial counterparts [25, 26].

## 2.5. Meteorological Data

Meteorological data from the PVsyst is used for simulation. It is also used as a reference for comparison and verification of the experimental results. The experimental setup consists of two pyranometers. One pyranometer is deployed horizontally and another is integrated with the S-N facial vertical bifacial module. Apart from that average monthly irradiance (GHI and DHI), temperature and rainfall patterns are also shown in **Fig. 4**.

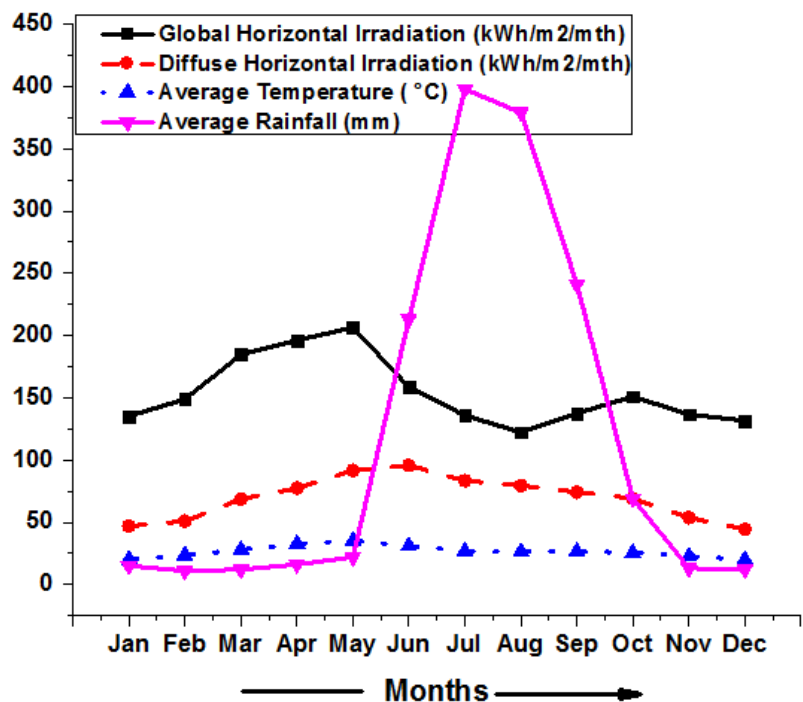

**Fig. 4:** *The average monthly irradiance, temperature and rainfall at experiment location (Raipur, Chhattisgarh)* [27]

## 2.6. Methodology

In the experimental setup, the bifacial PV modules are placed at about 81° tilt (near-vertical), one facing east and the other facing south. The monofacial PV modules are at a conventional tilt of 21° (equal to the latitude of the location) facing south. The nomenclatures of the PV modules are given in table 2 on the basis of technology/type and the inclination/orientation for ready reference. The pyranometers are deployed in such a way that one of them measures global horizontal irradiance (GHI), while the other measures plane of array (POA) irradiance of the front and rear-side for the bifacial PV module (south-facing). PV 200 records and stores I-V graphs. The solar survey 200R irradiance meter measures the typical plane of array (POA) irradiance and suction type temperature sensor measures the PV cell temperature and ambient temperature simultaneously while recording the I-V curve. The data is then transferred to a computer for further analysis. **Fig. 5** shows the complete experimental setup and methodology.

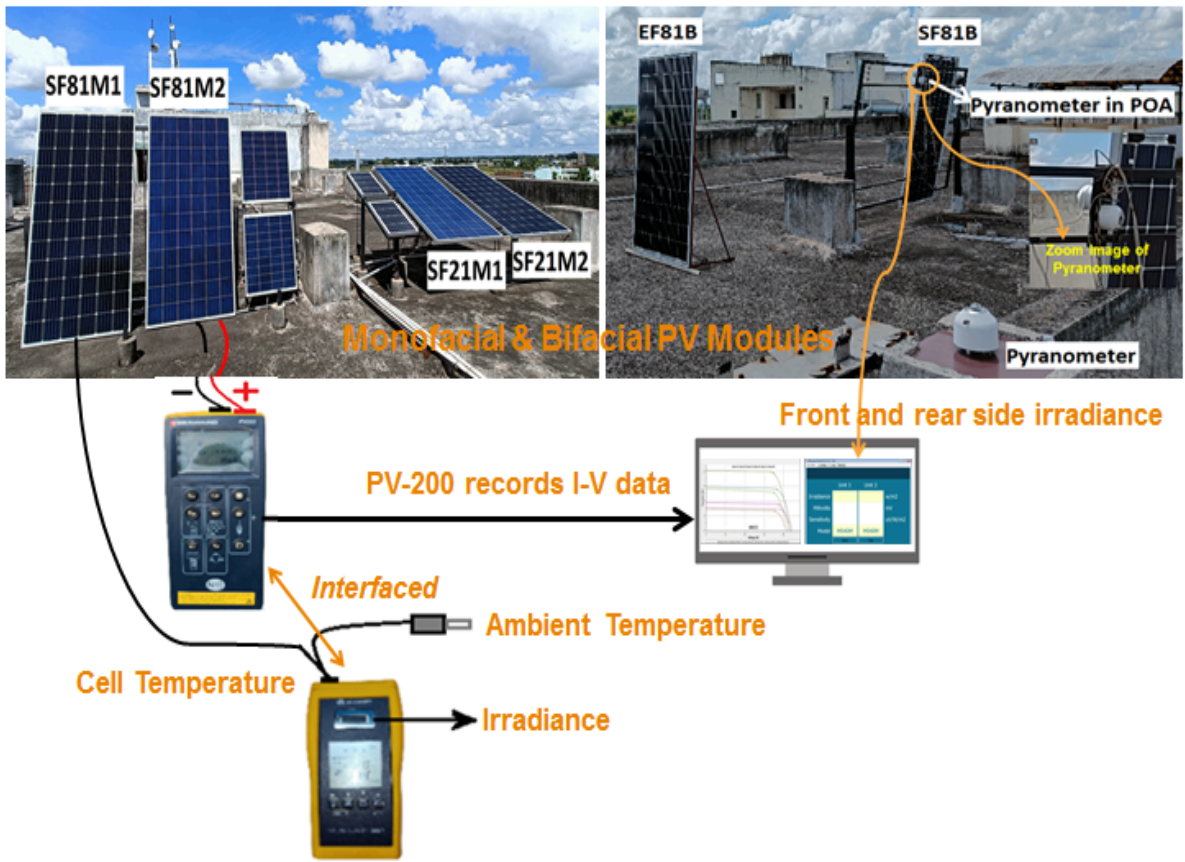

**Fig. 5:** *Complete outdoor experimental setup of monofacial and bifacial PV modules with measuring equipment*

I-V tracer (make: Seaward PV200) typically consists of a voltage source, a current meter, and a display and records 1000 no. of data. The voltage source is used to apply a range of voltage levels to the device under test (DUT), while the current meter measures the resulting current flowing through the DUT (Solar PV Module). The instrument plots the I-V curve of the DUT (Solar PV Module) on the display and records it for further analysis through a personal computer (PC).

Solar survey 200R irradiance meter with suction type temperature sensor is an integral part of the instrument. Irradiance meter used to measure the solar irradiance typically plane of array (POA) irradiance and suction type temperature sensor measures PV cell temperature and ambient temperature. The solar survey 200R irradiance meter typically consists of a sensor head and a handheld display unit. The sensor head contains a light sensor that measures the solar irradiance, while the display unit shows the measured values in real-time. The solar survey 200R irradiance meter is also equipped with a data logging feature with I-V tracer, which is connected wirelessly.

A pyranometer is used to measure the solar radiation and mounted at horizontal, tilt, and vertically as per applications. The pyranometer typically consists of a thermopile or photodiode sensor

(depending upon types of pyranometer), a glass dome to protect the sensor from the environment, and a signal processing unit to convert the sensor's output into a measurement of solar radiation. Specifications of pyranometer are**:** Class C Pyranometer ISO 9060 secondary standard, irradiance range up to 2000 W/m$^2$, response time < 0.7 s, zero offsets < 2 W/m$^2$.

The experiments have been carried out in two parts. The first set of experiments was conducted for four days from 22 May 2022 to 08 July 2022 with three PV modules: two bifacial and one monofacial between 09:15 AM - 01:45 PM. The second set of experiments included six PV modules, in which two bifacial and four monofacial PV modules were studied for several days from 08 October 2022 to 16 April 2023 between 07:15 AM - 04:45 PM. The day and the time in which the experiments were conducted are mentioned in **Table 3** section wise. We have collected 19-days data for a period of one year from May 2022 to April 2023, the important days like solstices and equinoxes are included in the experiments. We note here that due to resource and manpower limitations, as well as technical issues, the data for more days could not be recorded. As the experiments were performed manually, it was difficult to record data in summer and rainy seasons. So, representative days from each season were studied to obtain insight into the system behavior. The current-voltage (I-V) and power-voltage (P-V) curves were recorded for 19 days in the months from 22 May 2022 to 16 April 2023. Each I-V measurement has been done in the interval of 15 minutes or 30 minutes. Linear interpolation is used for calculating the energy for the entire day or experimental duration. The power curves for important days have been included in the results and discussion section. Calculation of energy generation is done as per the following relation.

$$\text{Energy/day} = \int_{t1}^{t2} p(t)dt = \sum_{t1}^{t2} p(t)dt \qquad (6)$$

where, p(t) = Instantaneous power

dt = Interval between the readings

**Table 3:** Experiment days and time detailed in Section I and Section II

| Section | Date | Duration of reading |
|---|---|---|
| Section I | 22/05/2022 to 08/07/2022 | 09:15 AM - 01:45 PM |
| Section II | 08/10/2022 to 16/04/2023 | 07:15 AM - 04:45 PM |

### 3. Results and discussion

The experiments are carried out in two phases, with section-specific dates and durations listed in **Table 3**. The irradiance profiles are given for the specific experiment dates. The corresponding output power is estimated from the experiments and compared with the PVsyst simulation. Later on, power output from the experiments were scaled to large PV plants and compared with PVsyst simulation.

## 3.1. Section I

In section-I, data and discussion for experiments carried out from 22 May 2022 to 08 July 2022 are presented. Four modules were deployed for this experiment, namely EF81B, SF81B, SF81MM and SF81PM. Nomenclature is mentioned in **Table 1** for reference. Current-voltage (I-V) measurements were recorded for every module. The data from these measurements were collected for a common time period between 09:15 AM and 01:45 PM. GHI profile is observed by pyranometers installed on the rooftop. The irradiance profile for 08 July 2022 shows significant variations. This day falls in the vicinity of rainy days. It shows that higher irradiance is expected due to intermittent availability of cloud coverage. And the atmosphere is less affected by dust, aerosol, and water vapor. July is the major monsoon month for Raipur and such variations are to be expected. The GHI profile is shown in **Fig. 6(a)** and the corresponding power curves for the EF81B, SF81B and SF81MM modules are shown in **Fig. 6(b)**.

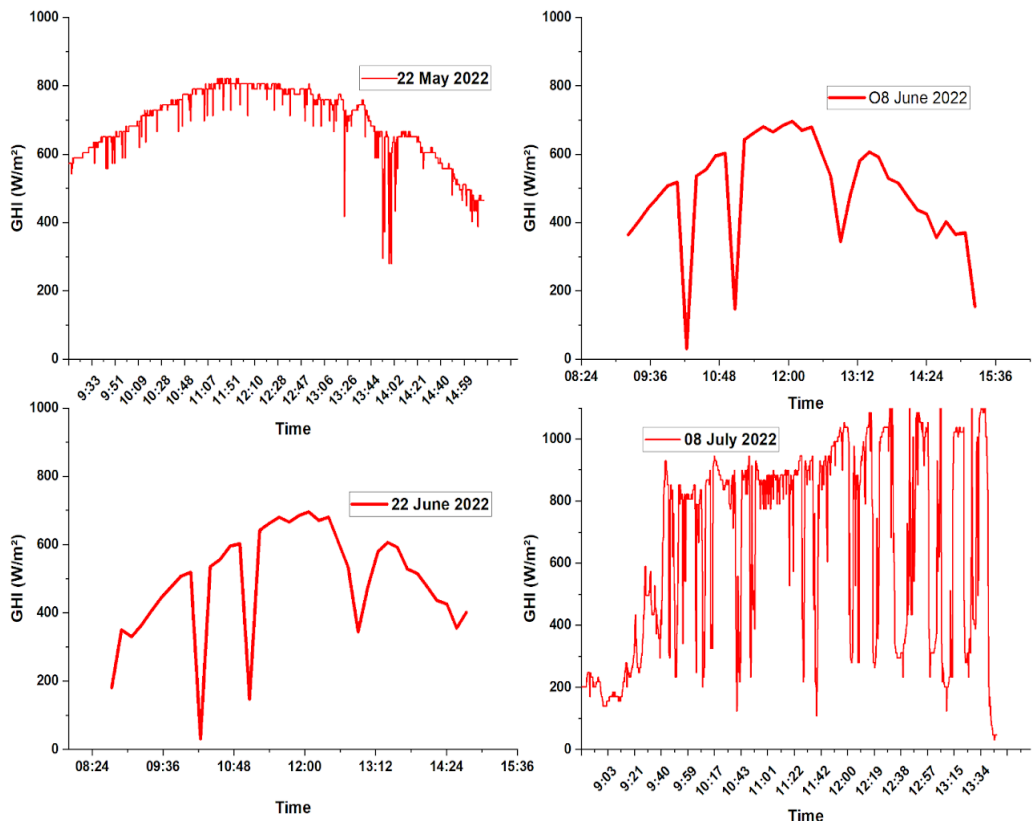

(a)

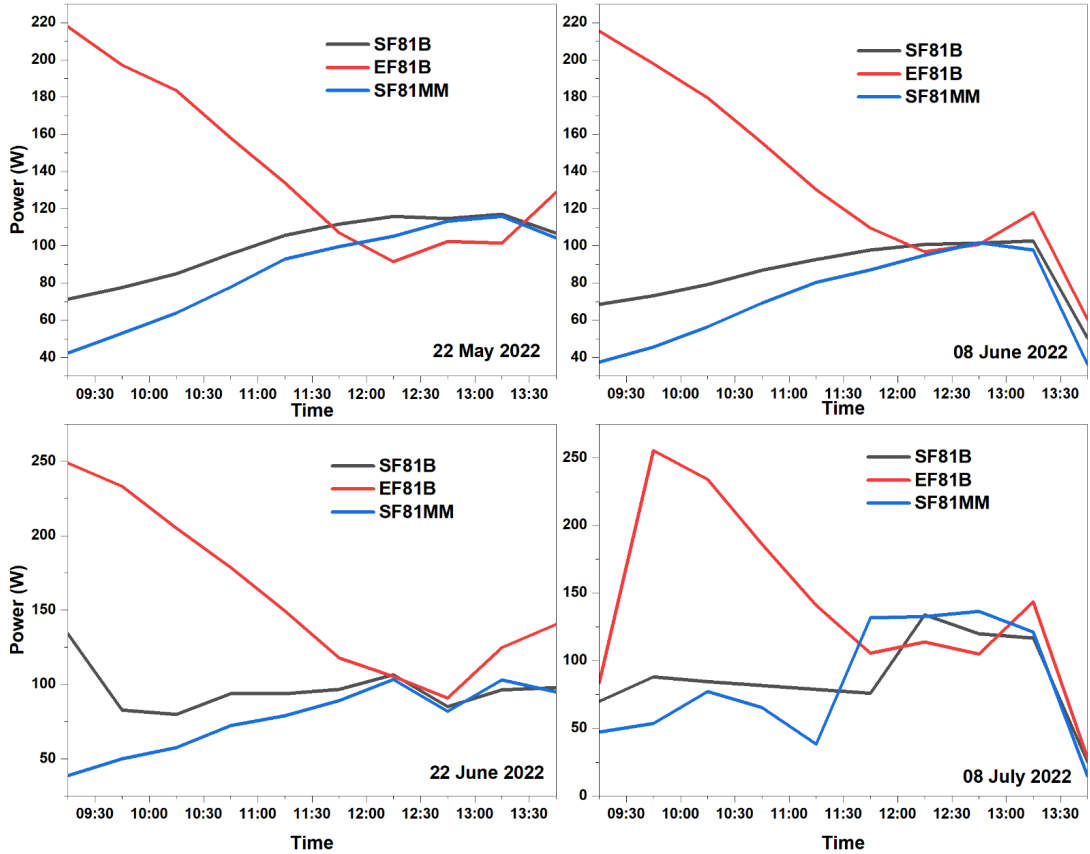

(b)

**Fig. 6:** *(a) GHI Profiles and (b) Power Curves recorded on experiment days in summer months*

Energy generation is calculated using **equation (6)** for discrete intervals. Energy generation for the East-West facing Bifacial PV module (EF81B), South-North facing Bifacial PV module (SF81B) and South-North facing monofacial (SF81MM) module has been shown in **Table 4**. The energy generation profile of the EF81B module with respect to SF81B and SF81MM is listed in **Table 5**. during the studied time frame (09:15 AM - 01:45 PM).

**Table 4:** PVsyst simulated results and the experimental results of the energy production (in kWh/kWp/day) of monofacial and bifacial PV modules on different dates from 09:15 AM to 01:45 PM.

| Experiment's Date | Average GHI (W/m$^2$) | Energy Generation (in kWh/kWp/day) | | | | | |
|---|---|---|---|---|---|---|---|
| | | EF81B | | SF81B | | SF81MM | |
| | | Exp. | PVsyst | Exp. | PVsyst | Exp. | PVsyst |
| 22-May-22 | 497 | 1.85 | 1.85 | 1.36 | 3 | 1.14 | 1.78 |
| 08-Jun-22 | 495 | 1.77 | 1.5 | 1.15 | 2.32 | 0.93 | 1.46 |
| 22-Jun-22 | 494 | 2.07 | 1.5 | 1.27 | 2.32 | 1.01 | 1.46 |
| 08-Jul-22 | 552 | 1.91 | 1.32 | 1.18 | 1.93 | 1.07 | 1.3 |
| **Average Output** | | **1.9** | **1.54** | **1.24** | **2.39** | **1.04** | **1.5** |

Table 5: Energy production of East-facing bifacial PV module (EF81B) with respect to South-facing bifacial module (SF81B) and South-facing monocrystalline monofacial module (SF81MM) for different dates from 09:15 AM to 01:45 PM.

| Date | Energy of EF81B / Energy of SF81B | Energy of EF81B / Energy of SF81MM |
|---|---|---|
| 22-05-22 | 1.42 | 1.63 |
| 08-06-22 | 1.59 | 1.92 |
| 22-06-22 | 1.65 | 2.07 |
| 08-07-22 | 1.59 | 1.7 |
| **Average** | **1.53** | **1.82** |

The results show that the vertical E-W facing Bifacial PV module gives two production peaks, one in the morning and one in the afternoon hours, instead of a single notable peak at midday provided by the monofacial module [28]. The minimum between the two power peaks of the bifacial module coincides with the maximum of the single peak of the monofacial modules. These two peaks coincide with high electricity demand in the morning and evening hours. It gives us a flexible energy profile according to the user's needs. This can be an extremely useful option for optimizing the energy generation profile. As a result, we can successfully optimize energy supply according to the demand if there are vertical oriented PV modules in E-W and S-N orientations. The energy generation of the E-W vertical bifacial PV module is 1.53 times more than the S-N vertical bifacial module, and 1.82 times more compared with the vertical south-facing monofacial module for a short time period (09:15 AM to 01:45 PM).

In this section, the data of conventional tilted modules, as well as the whole day energy generation profile, are not studied. To properly compare the conventional and vertical configurations, we present detailed analysis of six PV modules in different orientations in the next section.

### 3.2. Section II

In section-II, six PV modules, including two bifacial and four monofacial modules (EF81B, SF81B, SF81MM, SF81PM SF21MM, and SF21PM), were used to perform a comparative study. The full day experiment has been carried out for representative days between 08/10/2022 to 16/04/2023. And in order to make comparisons, full day reading of a standard time frame (07:15 AM - 04:45 PM) has been considered. This considers the whole day reading from near sunrise to sunset for the given location. Due to manpower and resource limitations, more data sets could not be recorded. Ideally, data for the whole year should be recorded. However, even this limited data provides useful insights into the efficacy of using vertical bifacial modules and a comparison with conventional tilted monofacial modules may be made. The power curves for important days have been included. These important days are solstices and equinoxes as listed in **Table 6**.

**Table 6:** Important days of the year (Solstice and Equinox)

| Date | Name |
|---|---|
| 22 June 2022 | Summer Solstice |
| 08 Oct 2022 | Near Autumnal Equinox* |
| 21 Dec 2022 | Winter Solstice |
| 23 March 2023 | Vernal Equinox |

* The autumnal equinox (22 September 2022) data could not be recorded due to equipment malfunction.

The solstices and equinoxes are astronomical phenomena that occur at key times in the Earth's annual orbit around the sun. The solstices occur twice a year, in June and December, and represent the year's longest and shortest days for the North hemisphere, respectively. The equinoxes, which occur twice a year in March and September, are also times when the lengths of day and night are equal. Experimental power curves for summer and winter solstices have been shown in **Figs. 7** and **8** respectively. Similarly, experimental power curves for near autumnal and vernal equinoxes have been shown in **Fig. 9** and **Fig. 10**.

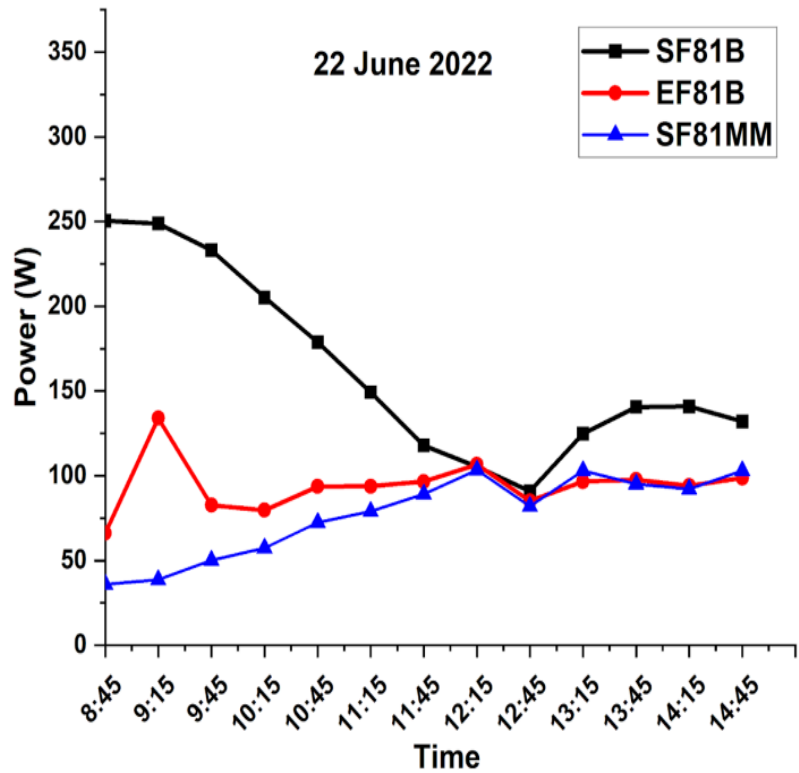

**Fig. 7:** *Experimental Power Curves for Summer Solstice (22 June 2022)*

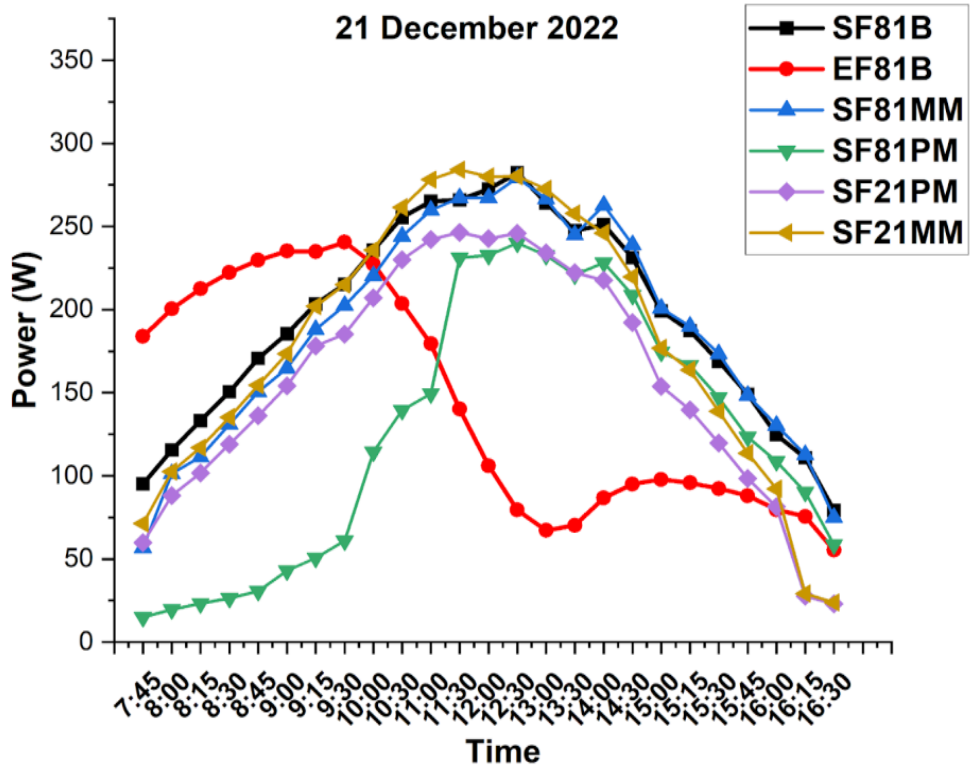

**Fig. 8:** *Experimental Power Curves for Winter Solstice (21 Dec 2022)*

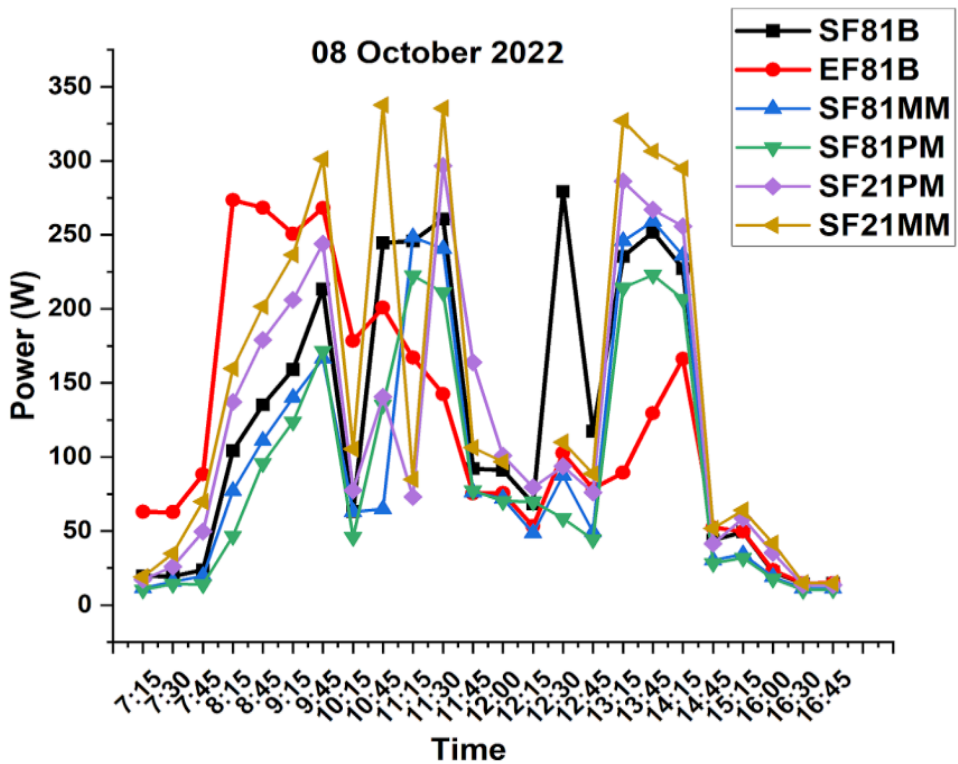

**Fig. 9:** *Experimental Power Curves for near Autumnal Equinox (08 Oct 2022)*

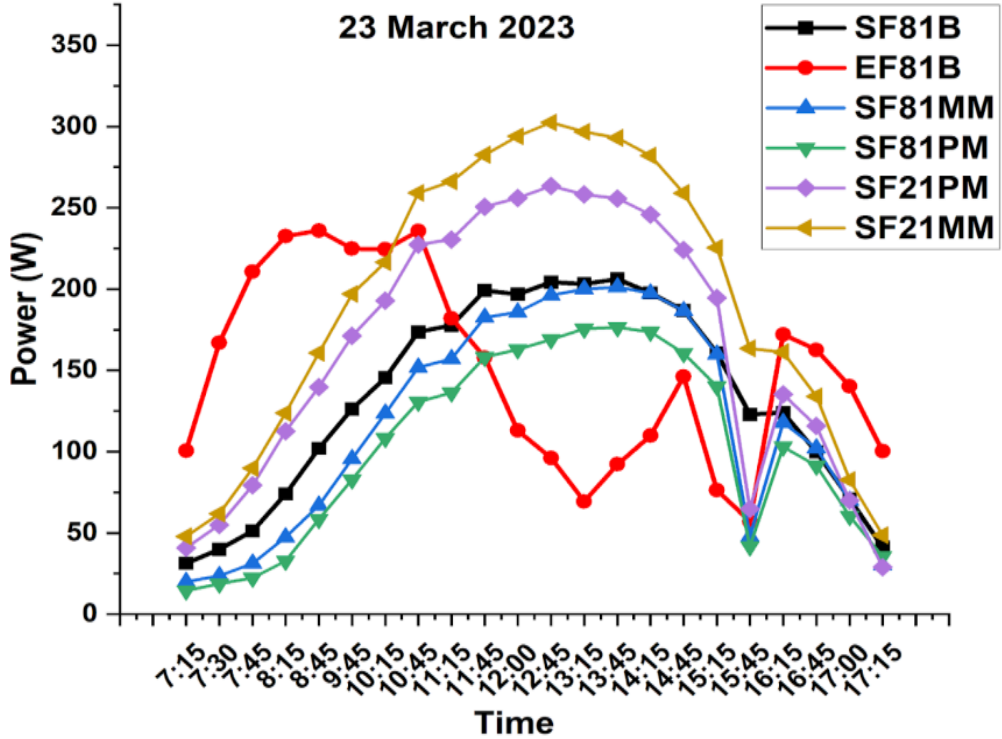

**Fig. 10:** *Experimental Power Curves for Vernal Equinox (23 March 2023)*

From the above figures, it can be readily seen that during the first two hours of the day, between 7:30 Am to 9:30 AM, the east-facing vertical bifacial PV module provides noticeably more power compared to all other configurations. And, as expected, the minimum of this module occurs at noontime, when south-facing modules are delivering their peak power. **Fig. 11** for the spring equinox also shows that the east-facing vertical bifacial module shows a power peak in the evening hours after 4 PM. As mentioned earlier, this dual-peak output of east-facing vertical configuration is desirable because it can provide a possible power balancing solution for the surge in electrical load during evening hours.

### 3.3. Correlation of experimental data with PVsyst simulations

The experimental results were also compared with PVsyst simulation. For this simulation, a trial version of PVsyst has been used. For simulation, six different PV plants are designed with the exactly same rating of the modules as per datasheet of the modules. PVsyst simulation has been carried out over a year. Performance Ratio (PR) is one of the metrics that allows trivial prediction of electrical energy yield (EY) of the PV plant [29]. In this section, PRs of six different PV modules are observed from PVsyst simulation over the year. This implies the methodology for getting predicted PR from the PVsyst simulation. Electrical energy yield can be written as mentioned in **equation (7)**.

$$EY = G_{POA} \times PR \times (\text{Rated Capacity}/G_{STC}) \qquad (7)$$

In the PVsyst, EY stands as energy injected into the grid (kWh), Rated Capacity stands as plant rated capacity/size (kWp), $G_{POA}$ stands for plane of array irradiance or global incidence (kWh/m$^2$), and $G_{STC}$ is 1000 W/m$^2$. From equation (7), PR can be written as:

$$PR = (EY \times G_{STC})/(G_{POA} \times Rated\ Capacity) \tag{8}$$

PRs are calculated for each PV module over a year as shown in **Fig. 11**. More practical design considerations can be seen in section 4.

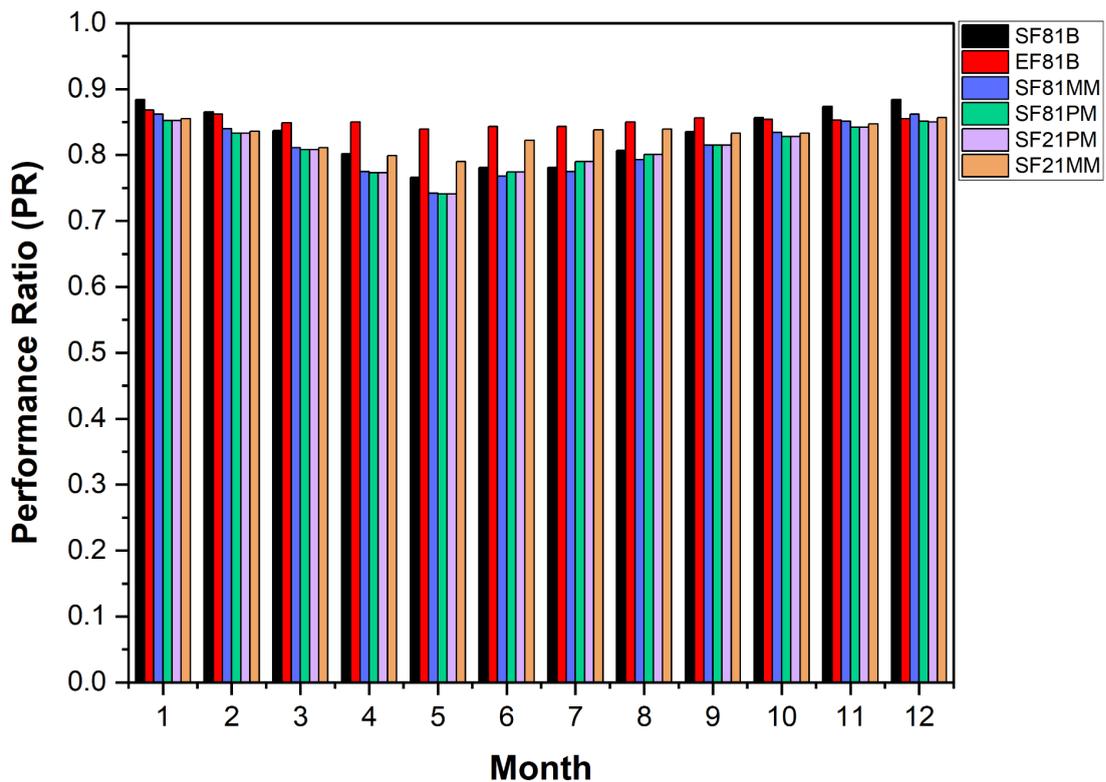

**Fig. 11:** *Performance ratio (PR) of different PV plants over the year from PVsyst simulation*

From the above figure, it can be seen that the EF81B PV module has consistent PR values over a year. While other plants have significant variations in the value of PR over a year. From March to September (regular 7 months), the PR for the EF81B PV module is greater than others. In the month of May, every PV plant encountered the lowest PR value while the EF81B PV plant showed a good PR value. In every case while designing the power plants, specific energy yields are considered for the different power plants. So that comparisons could be more realistic even after some minute change in the rated power capacity of the PV plants. Unlike the specific energy yield, the PR is not directly dependent on environmental factors and can be used to compare the system performance between installations at different locations and also in different orientations (as is the case here).

Experimental values and simulated values are given in **Table 7**, where experiment dates in the bold letters represent solstices and equinoxes. 05 Dec. 2022 was a cloudy day and is shown in italics. The energy generation profile of the EF81B module with respect to SF81B, SF81MM, and SF21MM is listed in **Table 8**. Apart from that SF81B is also compared with SF21MM during the studied time frame (07:15 AM - 04:45 PM). Overall, vertical bifacial modules are compared with conventional tilt modules as well as vertical mounting configurations.

**Table 7:** Experimental values of energy generation (kWh/kWp/day) by different PV modules measured in different seasons. Simulated values of energy generation (kWh/kWp/day) by monofacial/bifacial modules obtained from PVsyst are also given.

| Experiment Date | Average GHI (W/m²) | Energy Generation (in kWh/kWp/day) | | | | | | | | | |
|---|---|---|---|---|---|---|---|---|---|---|---|
| | | SF81B | | EF81B | | SF81MM | | SF81PM | SF21PM | SF21MM | |
| | | Exp. | PVsyst | Exp. | PVsyst | Exp. | PVsyst | | | Exp. | PVsyst |
| **08-Oct-22** | 449 | 3.57 | 3.16 | 3.54 | 2.08 | 2.75 | 3.08 | 2.83 | 3.86 | 4.12 | 4.16 |
| 23-Oct-22 | 365 | 4.88 | 3.16 | 3.85 | 2.08 | 4.56 | 3.08 | 4.35 | 5.32 | 5.49 | 4.16 |
| 07-Nov-22 | 408 | 3.98 | 3.78 | 3.01 | 1.88 | 3.58 | 3.69 | 3.46 | 4.11 | 4.19 | 4.17 |
| 20-Nov-22 | 426 | 4.39 | 3.78 | 3.34 | 1.88 | 3.94 | 3.69 | 3.76 | 4.03 | 4.14 | 4.17 |
| *05-Dec-22* | 369 | 3.17 | 4.36 | 2.51 | 1.84 | 2.91 | 4.27 | 2.77 | 2.94 | 2.99 | 4.38 |
| 13-Dec-22 | 366 | 4.13 | 4.36 | 2.86 | 1.84 | 3.7 | 4.27 | 3.58 | 3.98 | 4.11 | 4.38 |
| **21-Dec-22** | 396 | 5.19 | 4.36 | 3.31 | 1.84 | 4.86 | 4.27 | 3.91 | 4.67 | 4.78 | 4.38 |
| 14-Jan-23 | 340 | 4.16 | 4.37 | 2.89 | 2.02 | 3.53 | 4.28 | 3.54 | 3.87 | 3.97 | 4.54 |
| 23-Jan-23 | 375 | 5.03 | 4.37 | 3.56 | 2.02 | 4.61 | 4.28 | 4.45 | 4.8 | 4.92 | 4.54 |
| 08-Feb-23 | 469 | 5.27 | 3.83 | 4.31 | 2.23 | 4.72 | 3.73 | 4.79 | 5.41 | 5.59 | 4.66 |
| 22-Feb-23 | 415 | 4.06 | 3.83 | 4.07 | 2.23 | 3.42 | 3.73 | 3.27 | 4.45 | 4.58 | 4.66 |
| 11-Mar-23 | 445 | 3.62 | 3.23 | 3.94 | 2.63 | 3.78 | 2.91 | 3.06 | 4.46 | 4.61 | 4.82 |
| **23-Mar-23** | 510 | 4.13 | 3.23 | 4.57 | 2.63 | 3.49 | 2.91 | 3.39 | 5.53 | 5.78 | 4.82 |
| 11-Apr-23 | 516 | 2.88 | 2.52 | 4.29 | 3.11 | 2.4 | 2.43 | 2.38 | 4.6 | 4.75 | 4.97 |
| 16-Apr-23 | 483 | 2.79 | 2.52 | 3.85 | 3.11 | 2.36 | 2.43 | 2.36 | 4.62 | 4.79 | 4.97 |
| **Avg. Output** | | **4.1** | **3.65** | **3.6** | **2.22** | **3.6** | **3.53** | **3.46** | **4.44** | **4.6** | **4.51** |

Bold dates indicate equinoxes or solstices. Italics indicate a cloudy day. GHI was recorded using a pyranometer.

Generation profile of each PV plant has been extracted from the experiments at the given location. Later on, the generation (specific energy yield) profile is compared with the PVsyst simulation. Specific energy yield in **Table 7** is obtained from experiments as well as simulations. Specific energy yield can be written as

$$\text{Specific energy yield} = \frac{Yearly\ energy\ generation\ (kWh)}{Rated\ plant\ capacity\ (kWp)\ x\ 365} \tag{9}$$

From Table **7**, it can be clearly seen that there's a significant difference between experimental and simulation results. Vertical setup is explored in detail for this analysis along with conventional setup. Experimental and simulated results are also compared. As seen in **Fig. 12**, simulation for the East facing bifacial module (EF81B) shows a significant mismatch with the experimental data. It indicates that PVsyst default parameters are not providing a realistic picture of the east-facing vertical bifacial setup and are providing a huge underestimate of the power output. Some of the possible reasons for this discrepancy may be soiling factor overestimation for the vertical modules, meteorological data and irradiance profile mismatch for the studied location.

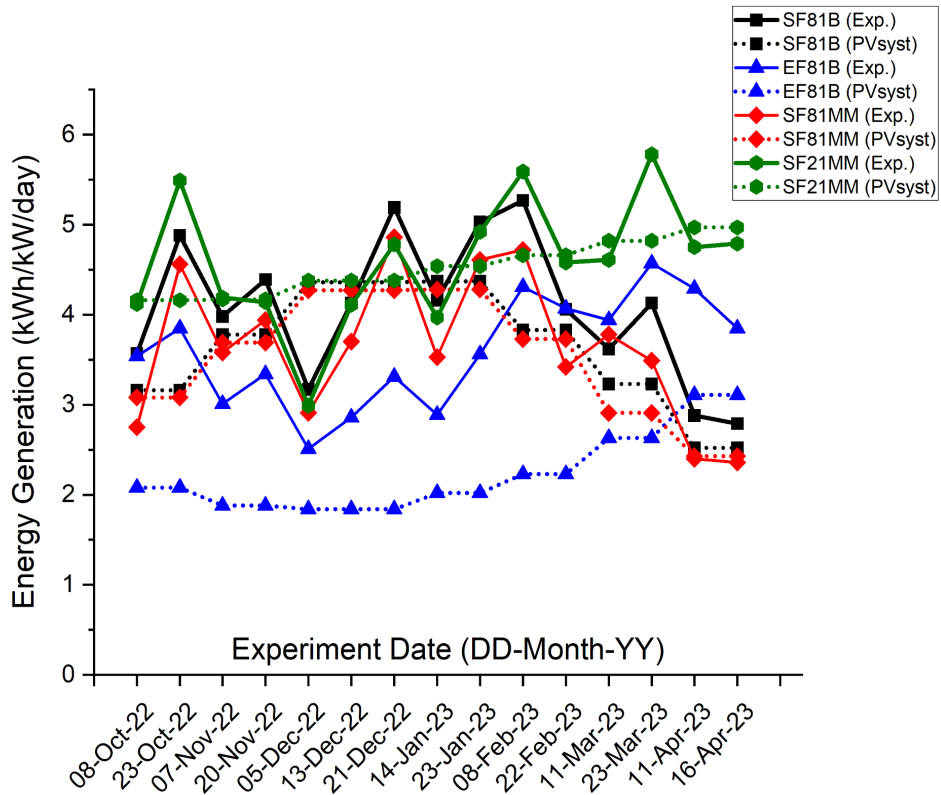

**Fig. 12:** *Specific energy yield for SF81B, EF81B, SF81MM, and SF21MM from simulations and experiments*

From experimental results, it can be observed that EF81B gives more energy generation in the months of February, March, and April. And performance is lesser in the months of December and January. From 22/05/2022 to 08/07/2022 during the studied time frame of 09:15 AM - 01:45 PM, energy generated by EF81B is 1.53 times more than the SF81B, and 1.82 times more than SF81MM. Overall from 08/10/2022 to 16/04/2023 during the studied time frame of 07:15 AM - 04:45 PM, energy generated by EF81B is 0.91 times of SF81B, and 1.04 times more than SF81MM. Although more experimental data are needed. But, it's sufficient to identify the trend. Vertical bifacial modules have the ability to generate substantial power while occupying only a small portion of the space required by traditional tilted monofacial modules.

**Table 8:** Energy production of EF81B with respect to SF81B and SF81MM on different dates from 07:15 AM - 04:45 PM

| | Ratio of Energy Generation (in kWh/kWp/day) | | | |
|---|---|---|---|---|
| Date | EF81B / SF81B | EF81B / SF81MM | EF81B/SF21MM | SF81B/SF21MM |
| **08-10-22** | 0.99 | 1.29 | 0.86 | 0.87 |
| 23-10-22 | 0.79 | 0.84 | 0.70 | 0.89 |
| 07-11-22 | 0.76 | 0.84 | 0.72 | 0.95 |
| 20-11-22 | 0.76 | 0.85 | 0.81 | 1.06 |
| *05-12-22* | *0.79* | *0.86* | *0.84* | *1.06* |
| 13-12-22 | 0.69 | 0.77 | 0.70 | 1.00 |
| **21-12-22** | 0.64 | 0.68 | 0.69 | 1.09 |
| 14-01-23 | 0.69 | 0.82 | 0.73 | 1.05 |
| 23-01-23 | 0.71 | 0.77 | 0.72 | 1.02 |
| 08-02-23 | 0.82 | 0.91 | 0.77 | 0.94 |
| 22-02-23 | 1.00 | 1.19 | 0.89 | 0.89 |
| 11-03-23 | 1.09 | 1.04 | 0.85 | 0.79 |
| **23-03-23** | 1.11 | 1.31 | 0.79 | 0.71 |
| 11-04-23 | 1.49 | 1.79 | 0.90 | 0.61 |
| 16-04-23 | 1.38 | 1.63 | 0.80 | 0.58 |
| **Average** | **0.91** | **1.04** | **0.79** | **0.90** |

A comparative analysis of different mounting configurations including the vertical and conventional setup is explored. The experiments from 08/10/2022 to 16/04/2023 during the studied time frame of 07:15 AM - 04:45 PM shows a quantified as well qualitative observation for the central-east Indian location. Overall, EF81B gives more generation than SF81MM whereas EF81B gives slightly lower energy generation than SF81B. The energy generated by EF81B is 0.91 times of SF81B, 1.04 times more than SF81MM, and 0.79 times with respect to SF21MM. Even energy generation of SF81B is 0.9 times of SF21MM.

SF81B gives more generation than SF21MM from October to February. After February, SF81B gives lower generation than SF21MM till October. Whereas EF81B gives more generation than SF81B and SF81MM from February to early week of October. For the same duration, from February to the early week of October, EF81B performed best compared to other months. The generation of each experimental day is verified with PVsyst simulation. For conventional azimuth and the corresponding tilts, PVsyst results are comparable and match the experimental conditions. The simulation result finds that PVsyst underestimates the generation of the east facing vertical modules. Default values such as the incidence angle modifier (IAM) factor, PV loss due to irradiance and temperature level for this azimuth at various tilts needed to be modified. Energy gain is comparable or not much affected with vertical setup. Additionally, we get a usable land without harnessing much energy. It enables the PV plant to serve a dual-purpose for land use.

## 4. Grid Tied Solar PV Plant Design for One Acre Area

Optimum design plays a crucial role while designing the PV power plants. The size of the PV power plants always disappoints when energy demand is not met. This might be due to the improper design and improper mounting configurations. Design constraints stand for specific tilt, row spacing, effective area and nature of the land, such as whether it is plain, undulating, or sloped. Eventually, improper PV plant design leads to the wastage of land as well as delays in recovering of the initial cost of the plant. As a consequence, the system experiences a longer payback period.

Newer technology modules with proper design can meet the high load demands using the same area. Some of the well-known existing PV power plants in India are given in **Table 9**. One can observe the installed capacity per acre of land and see that installations per unit acre aren't consistent.

**Table 9**: PV plant capacity per acre area [30]

| Serial No. | Name of the Plant (Year commissioned) | Plant Capacity (kWp) | Area (acre) | kWp/acre |
|---|---|---|---|---|
| 1 | Vankal Solar Park, Mizoram, India (2023) | $2 \times 10^4$ | 194 | 103.09 |
| 2 | Bhadla Solar Park, Rajasthan, India (2020) | $2.25 \times 10^6$ | 14000 | 160.36 |
| 3 | Pavagada Solar Park, Karnataka, India (2019) | $2.05 \times 10^6$ | 13000 | 157.69 |
| 4 | Kurnool Ultra Mega Solar Park, Andhra Pradesh, India (2017) | $1.00 \times 10^6$ | 5932 | 168.58 |
| 5 | Welspun Solar MP project, Madhya Pradesh, India (2014) | $1.51 \times 10^5$ | 750 | 201.33 |

For the present work, monocrystalline monofacial PV modules (375 Wp), polycrystalline (330 Wp) and the n-type PERT bifacial PV modules (355 Wp) have been simulated in PVsyst. Apart from that experimental results are also scaled up for one acre area.

**4.1. Inter-row spacing of PV array**

For conventional PV plants, inter-row spacing is primarily determined by the worst-case shading conditions that generally occurs during the winter solstice. This can be easily seen from sun-path diagram of the given location. The worst-case scenario can be calculated as shown in the **Fig. 13**.

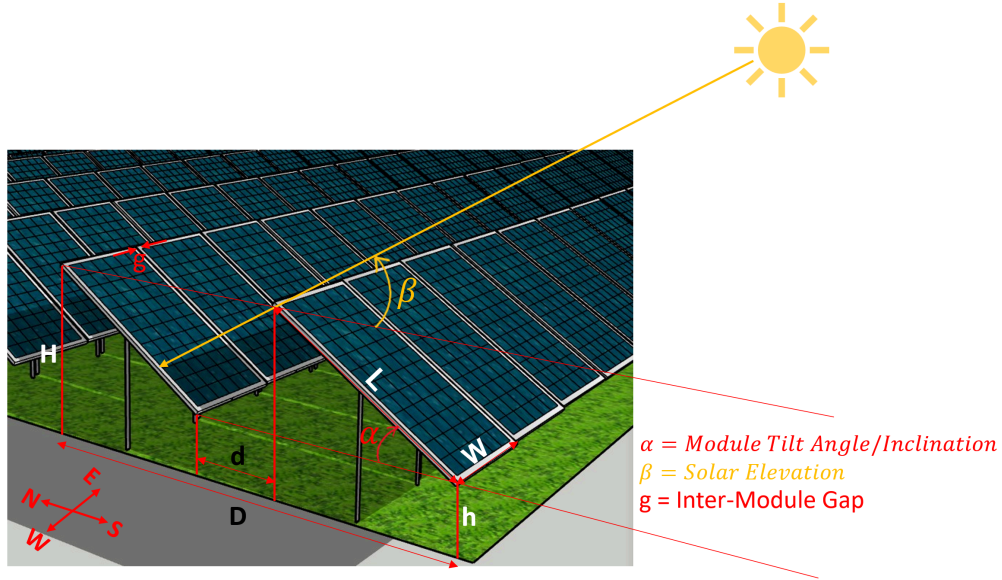

**Fig. 13:** *Inter-row spacing of PV array*

For the worst-case shading scenario, sun-path of December 22[nd] is considered. Assuming the sun azimuth to be 48 degrees and sun elevation angle of 27 degrees, the inter-row spacing can be determined by the following equation.

$$d = \ \cdot \frac{\cos\cos{(sun\ azimuth)}}{\tan\tan} \tag{10}$$

For the given location, module tilt is 23 degrees. So, from equation 10, calculated inter-row spacing is 1.026 m. This inter-row spacing is for the minimum clearances. Assuming the upper height (H) as 0.92 m and lower height (h) as 0.15 m from ground. If a table of two strings are considered. Then in this case minimum inter-row space will be doubled. So, minimum inter-row space will be 2.052 m. Similarly for the vertical PV plants, the spacing between the modules is considered as 3 meters. Considering a 0.5-meter clearance on each side and average width of the farming tractor as 1.85 meter, the optimum distance comes out to be 2.85 meter (~3 meters).

**4.2. Calculation of Plant Sizes for Different Configurations on 1 Acre of Land**

After accounting for the minimum inter-row spacing, the PV modules can be effectively distributed across the 1-acre area (63.61 m × 63.61 m) to maximize land utilization and system capacity. Using one acre as a direct input for active module area often results in an overestimated plant capacity, since it does not account for inter-row spacing, walkways, and other non-module areas. To address this, both conventional (tilted) and vertical configurations are geometrically distributed to more accurately reflect real-world layout constraints. It gives the geometrical upper limit of the plant size (high module density installation) assuming the land is flat.

Three kinds of grid tied PV systems have been simulated. The first one is a conventional plant (south facing monofacial modules at conventional tilt) and two different vertical PV plants (S-N and E-W facing bifacial modules). Since a hypothetical 1-acre PV plant was simulated, we relied on the default values by the PVsyst. Apart from that, specific energy yields are compared with experimental results, and average daily generation is assumed to remain constant throughout a given month in the PVsyst simulation. Based on these assumptions from the simulations, the study is carried out. Design of grid tied solar PV plants and energy generation values for 1 acre are shown in **Fig. 14**. For conventional PV plant, calculated inter-row spacing is 1.026 m and inter-module space is assumed as 3 mm. It populates the 63 modules along one of the edges of the land and by considering inter-row spacing, 22 modules can be fit along other edges. So, it populates 1386 (63 × 22) modules for conventional PV plant. It results the system size of 519.75 kWp system (1386 × 375 Wp). Similarly for the poly crystalline module with 330 Wp capacity results the system size of 457.38 kWp. Same exercise is repeated for the vertical PV plant. In this case, inter-row space of 3 m is assumed and module is mounted in landscape mode. Along one edge, 31 modules can be accommodated. The inter-row space is assumed as 3 m and mounting structure width is assumed as 0.3 m. Then it populates 20 modules on another edge. So, it populates 620 (31 × 20) modules in 1-acre area resulting in a PV plant size of 220.1 kWp (620 × 355 Wp).

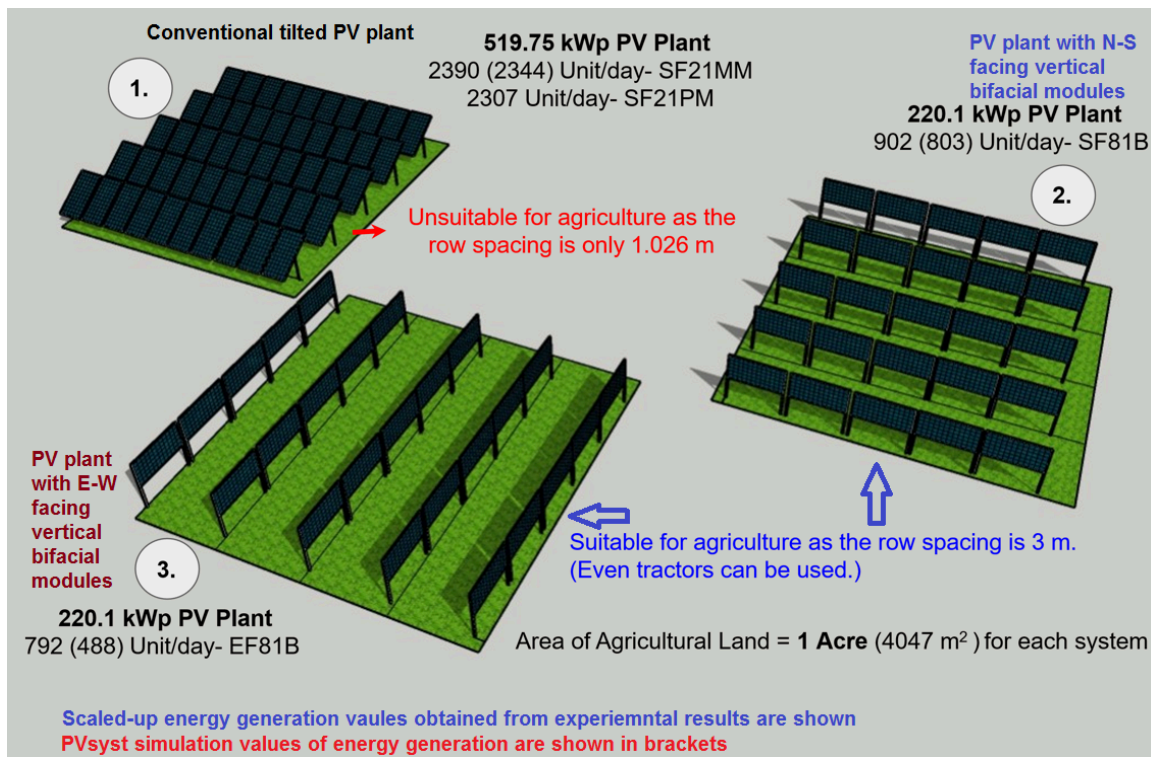

**Fig. 14:** *Grid tied solar PV plant design for 1 acre area for the studied location (Raipur, Chhattisgarh)*

From **Fig. 14**, we can see that the south-facing vertical bifacial PV plant generates about ~38% of the energy provided by the conventional tilted plant. Likewise, the east-facing vertical bifacial plant provides about ~32% of the energy generated by the conventional plant. These values are quite promising considering the fact that the same land can be used for agricultural purposes. We note here that in section 3.3, the simulations provide an underestimate of the actual power generation for the vertical bifacial modules. So, the results of section 4 should also be taken as an underestimation and not an upper bound. Vertical installations need to be studied in more detail and simulation models need to be updated to get a more realistic picture.

Diverse locations across India from higher to lower latitudes can give a more comprehensive understanding of how latitude and climatic conditions affect PV energy generation. Three locations have been considered as shown in **Table 10,** namely Leh (Cold and mountainous), Raipur (Tropical wet and dry climate) and Palakkad (Tropical monsoon and humid). In this table, energy generated per year per kWp of installed capacity is shown for comparison.

Table 10: Specific energy yield (kWh/kWp/year) per acre area for different locations

| Serial No. | Location (Latitude in degrees North) | Conventional (kWh/kWp/yr) | SF81B (kWh/kWp/yr) | EF81B (kWh/kWp/yr) |
|---|---|---|---|---|
| 1 | Leh, Ladakh, India (34.16°) | 1883 | 1388 | 1167 |
| 2 | Raipur, Chhattisgarh, India (21.16°) | 1524 | 1022 | 907 |
| 3 | Palakkad, Kerala, India (10.77°) | 1539 | 878 | 901 |

Palakkad, located close to the equator, represents a low-latitude site, which would typically be expected to receive higher solar irradiance throughout the year. Despite its lower latitude and theoretically higher solar potential, Palakkad exhibits the lowest specific energy yield for vertical SF81B and EF81B system configurations, though it's conventional system yield is comparable to that of Raipur (21.16°). The local environmental conditions such as persistent cloud cover, humidity and lower albedo affect the performance of system. On the other hand, Leh has higher solar insolation, low temperature and higher albedo, consistently records the highest energy yields across all system configurations. The specific energy yield ratio of S-N vertical configuration to conventional configuration is an impressive 73.7% for Leh. It implies that the higher latitude sites are even more favourable for vertical PV installations.

### 4.3. Implications for Agrivoltaics in Indian subcontinent

The energy generation is a bonus for the farmer who doesn't have to worry about his land being semi-permanently occupied by the PV plant. In many scenarios, it is even more appealing for farmers, where many of them live inside or very near to their farmland. This dual-use of land would allow them to produce their own electricity, with a possibility of selling it to the grid as well. Some of the agrivoltaic setups have resorted to the use of elevated metal structures for mounting the PV modules [31]. It adds to the cost of the PV setup and is difficult to clean and maintain. The simple arrangement simulated here for 1-acre land, is easy to implement and maintenance is also relatively easy. Additionally, sprinkling systems widely employed by farmers in India can act as an effective cleaning mechanism for the vertical bifacial PV modules installed in the simple manner as shown in **Fig. 14**. This ease of cleaning and maintenance is not available in overhead agrivoltaic systems. Vertical bifacial modules have the ability to generate substantial power while occupying only a small portion of the space required by traditional tilted monofacial or bifacial modules. As a result, they are highly suitable options for noise barriers, AgriPV (agriculture-integrated photovoltaic) [32], BIPV (building-integrated photovoltaic), and VIPV (vehicle-integrated photovoltaic) applications. Other dual land-use applications may be in animal husbandry, where pasture lands may utilize vertical modules without affecting the grazing area. Plant nurseries, greenhouses and educational or residential campuses may also leverage the benefits of vertical bifacial modules.

## 5. Conclusion and future scope

This research work evaluates the effectiveness of an E-W facing vertically oriented bifacial solar photovoltaic (BPV) module and compares it to a S-N facing vertically oriented bifacial module. The bifacial modules were also compared with S-N facing mono facial PV modules. Six modules including two bifacial modules were used to perform a comparative study. Current-voltage (I-V) and power-voltage (P-V) curves were recorded for 19 days in the months of May 2022 to April 2023. Two sets of experiments were carried out. First, from 22 May 2022 to 08 July 2022, and the second set was carried out from 08 Oct 2022 to 16 April 2023. For the first set of experiment, it has been observed that the East-facing bifacial PV module provided 1.53 times more energy generation with respect to the vertical South-facing bifacial module, and 1.82 times more with respect to the vertical South-facing monofacial module during the studied time period (09:15 AM to 01:45 PM). For the second set of experiments, it has been observed that energy generation by EF81B is 0.91 times of SF81B, and 1.02 times more than SF81MM during the studied time frame of 07:15 AM - 04:45 PM. For the months of February to April, generation of East-West facing bifacial is more compared to south facing bifacial and south facing monocrystalline monofacial modules. PVsyst simulations showed relatively lower energy generation for vertical bifacial modules compared to experimental values, highlighting the need for improved simulation models that can account for accurate for daily as well as partial-day energy generation. As per experimental results, bifacial PV modules are highly suitable for dual land-use applications. They enable simultaneous electricity generation while allowing the same land to be utilized for additional purposes.

**CRediT authorship contribution statement**

**Nishant Kumar:** Writing – original draft, Visualization, Software, Methodology, Formal analysis, Data curation.
**Shravan kumar Singh:** Writing – review & editing, Formal analysis, Visualization.
**Nikhil Chander:** Writing – review & editing, Project administration, Methodology, Investigation, Formal analysis, Conceptualization.

**Declaration of competing interest**

The authors declare that they have no known competing financial interests or personal relationships that could have appeared to influence the work reported in this paper.

**Data availability**

All the relevant data are provided in the manuscript.